\documentclass{PoS}

\title{Physics case of the very high energy electron--proton collider, VHEeP}

\ShortTitle{Very high energy electron--proton collider, VHEeP}

\author{A. Caldwell\\
        Max Planck Institute for Physics, Munich\\
        E-mail: \email{caldwell@mpp.mpg.de}}

\author{F. Keeble\\
        University College London\\
        E-mail: \email{fearghus.keeble.11@ucl.ac.uk}}

\author{E. Simpson Dore\\
        University College London\\
        E-mail: \email{emma.dore.13@ucl.ac.uk}}

\author{\speaker{M. Wing}\thanks{Also at DESY.}\\
        University College London\\
        E-mail: \email{m.wing@ucl.ac.uk}}

\abstract{The possibility of a very high energy electron-proton (VHEeP) collider with a
centre-of-mass energy of 9 TeV has been presented at previous workshops.  These 
proceedings briefly summarise the VHEeP concept, which was recently published, and 
developments since then, as well as future directions.  At the VHEeP collider, with a
centre-of-mass energy 30 times greater than HERA, parton momentum fractions, $x$,
down to about $10^{-8}$ are accessible for photon virtualities, $Q^2$, of 1 GeV$^2$.  This 
extension in the kinematic range to low $x$ complements proposals for other 
electron--proton or electron--ion colliders.}

\FullConference{XXV International Workshop on Deep-Inelastic Scattering and Related Subjects\\
		3-7 April 2017\\
		University of Birmingham, UK}

\begin{document}

\section{Introduction}

Simulations of proton-driven plasma wakefield acceleration~\cite{pdpwa} have shown that acceleration of electrons 
to TeV scales over km scales is possible.  The AWAKE experiment~\cite{awake} at CERN is investigating and 
measuring the effect for the first time.  Assuming the current and future investigations at AWAKE are successful, possible 
applications of the AWAKE acceleration scheme are being considered in which an electron beam is accelerated to high 
energies over short distances.  Making strong use of CERN infrastructure, a very high energy electron--proton (VHEeP) 
collider~\cite{vheep} has been proposed with an electron energy of $E_e = 3$\,TeV and $E_p =7$\,TeV, leading to a 
centre-of-mass energy of about 9\,TeV.

The proposed scheme for VHEeP leads to a baseline centre-of-mass energy a factor of 30 times higher than HERA and 
a corresponding extension to low Bjorken $x$ of a factor of 1\,000, with a similar extension to high photon virtuality, $Q^2$, 
depending on the luminosity.  The layout for VHEeP leads to modest luminosities, with an integrated luminosity during the 
collider's lifetime of $10 - 100$\,pb$^{-1}$.  Studies showed that the nature of the strong force and its explanation within 
QCD is not well understood at values of Bjorken $x$ down to $10^{-8}$, corresponding to $Q^2$ of about 1 GeV$^2$.  It is expected that 
effects of saturation of the structure of the proton and hadronic cross sections will be observed given the unphysical nature 
of various models when extrapolated to VHEeP energies.  This can be studied in inclusive deep inelastic scattering events, 
by measuring the total photon--proton cross section, as well as tagging vector mesons.  Given the rise of the cross sections 
to low Bjorken $x$, even with modest luminosities, numerous events are expected in all of these reactions.  Exotic, 
high-$Q^2$ physics can also be studied at VHEeP, with particular sensitivity to leptoquark production up to the kinematic 
limit of 9\,TeV, as well as quark substructure.

The physics case of VHEeP developed in the original publication~\cite{vheep} was briefly described in the previous paragraph,  
with the rest of the proceedings concentrating on developments since.  In Section~\ref{sec:recent}, recent developments 
are discussed, in particular the use of modern Monte Carlo tools and comparison with the previous studies.  In 
Section~\ref{sec:workshop}, a workshop dedicated to the particle physics case for VHEeP is briefly summarised.  The 
proceedings are concluded in Section~\ref{sec:summary}.


\section{Recent studies and Monte Carlo development}
\label{sec:recent}

Studies published previously~\cite{vheep} used the {\sc Ariadne}~\cite{ariadne} Monte Carlo programme to investigate 
$ep$ collisions at 9\,TeV.  The {\sc Ariadne} Monte Carlo programme is written in Fortran and was used standalone and so a 
more modern framework and programmes were investigated.  The {\sc Ariadne} predictions were also made using the 
CTEQ2L~\cite{cteq2l}  proton parton density functions (PDFs); more modern Monte Carlo programmes are interfaced to 
LHAPDF~\cite{lhapdf} and so allow the use of more recent PDFs.  The {\sc Rivet} toolkit (Robust Independent Validation of 
Experiment and Theory)~\cite{rivet} is a system for validation of Monte Carlo generators and so once a routine has been 
developed to investigate VHEeP collisions, then several Monte Carlo programmes can be used; the Monte Carlo programmes 
can also be compared to existing data and so their model(s) validated.

\begin{figure}
\begin{center}
\includegraphics[width=0.46\textwidth]{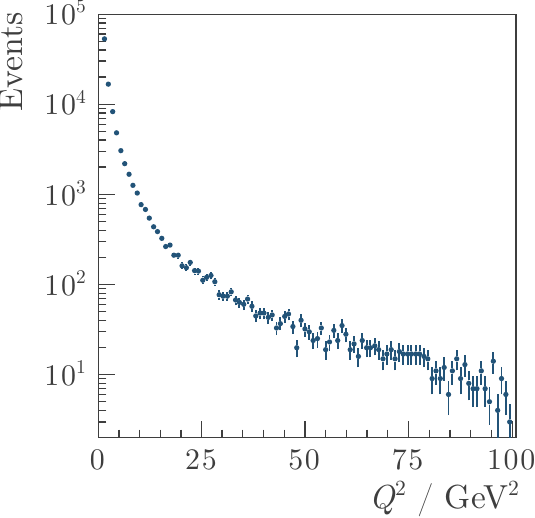}
\put(-30,150){\makebox(0,0)[tl]{(a)}}
\put(195,150){\makebox(0,0)[tl]{(b)}}
\includegraphics[width=0.52\textwidth]{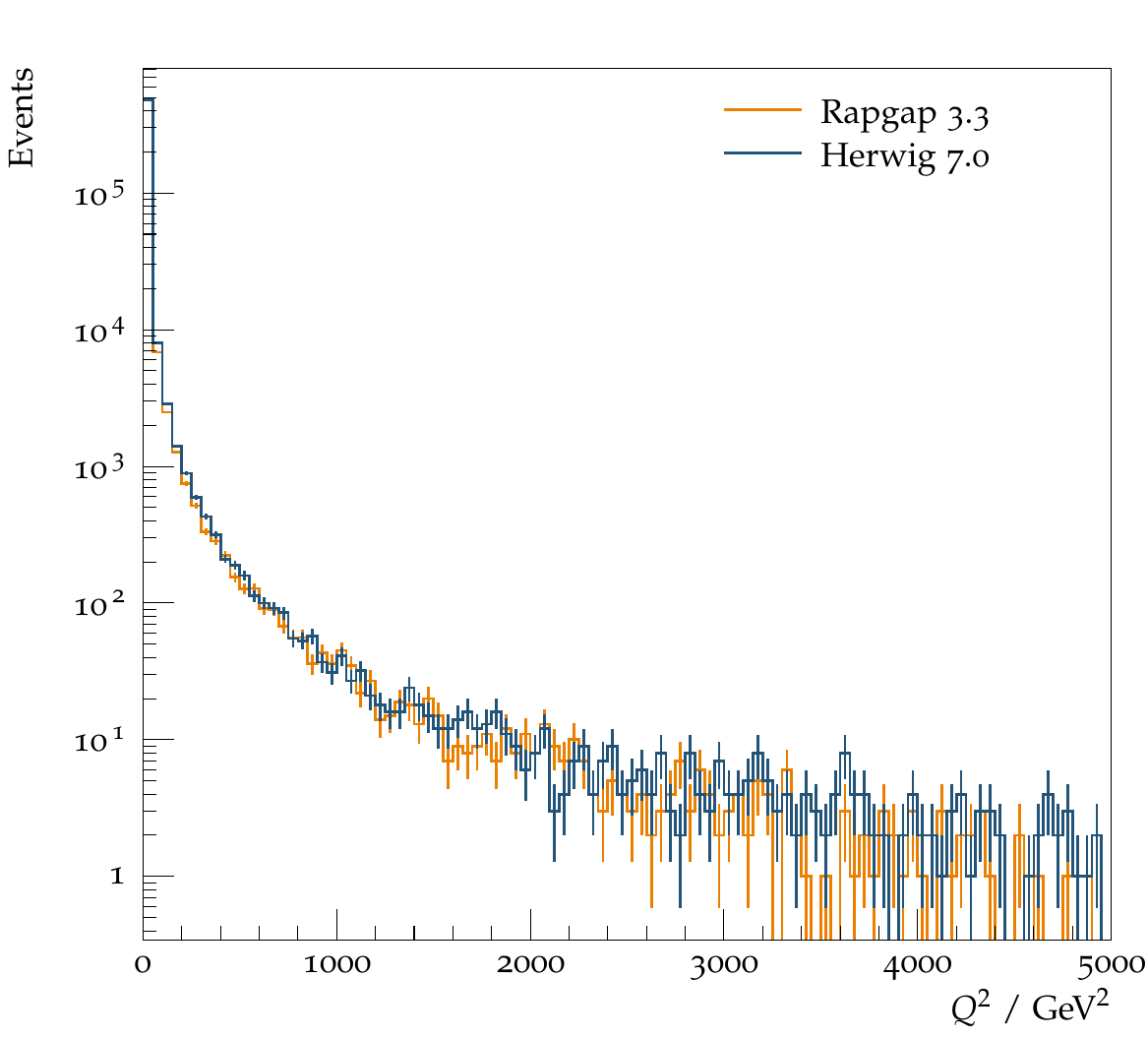}
\includegraphics[width=0.46\textwidth]{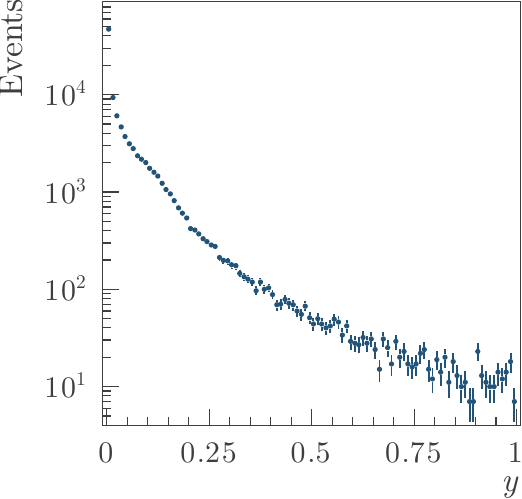}
\put(-30,150){\makebox(0,0)[tl]{(c)}}
\put(195,150){\makebox(0,0)[tl]{(d)}}
\includegraphics[width=0.52\textwidth]{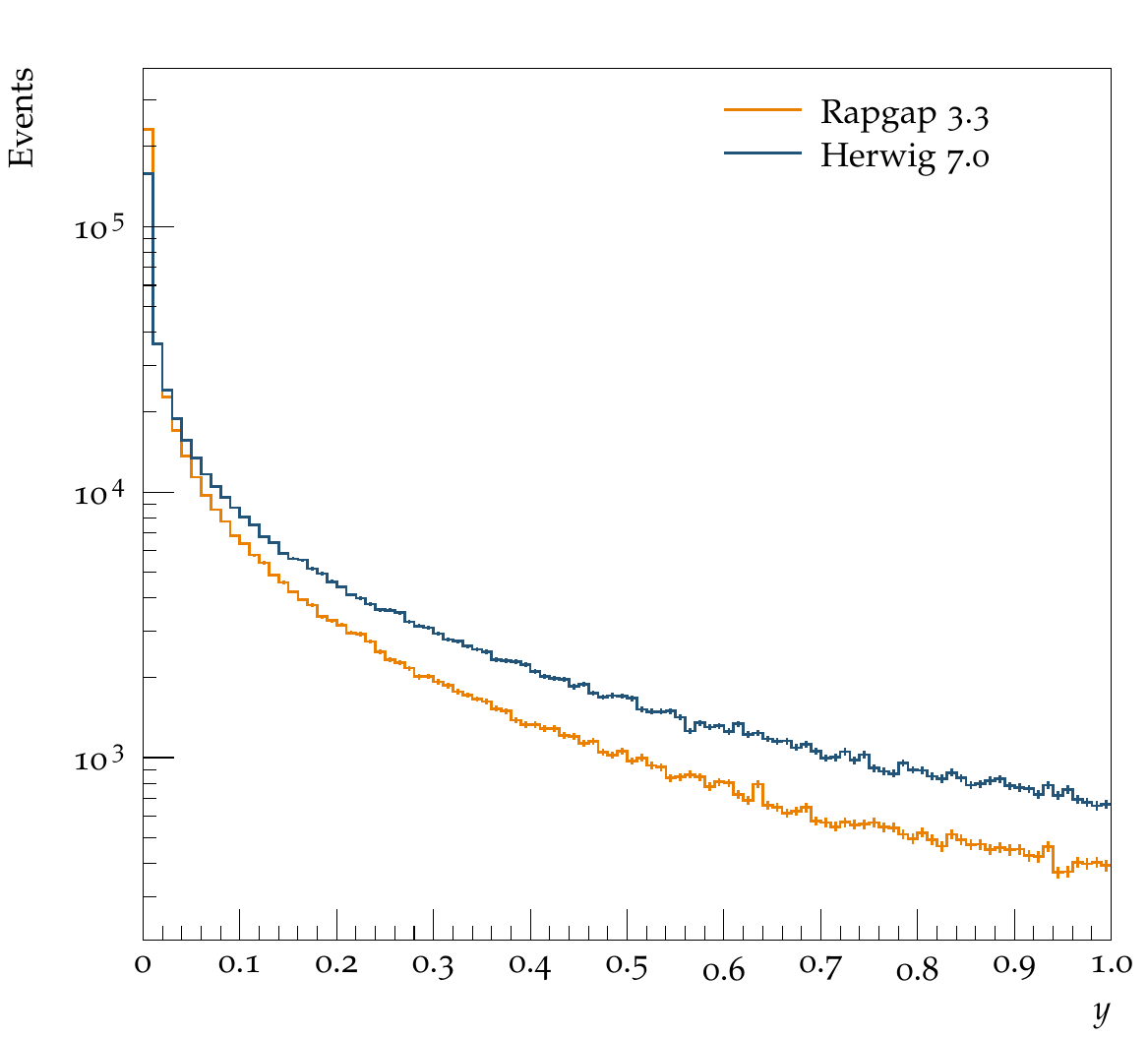}
\end{center}
\caption{Kinematic distributions for (a,b) photon virtuality, $Q^2$, and (c,d) inelasticity, $y$, for a small sample of events at VHEeP.  (a,c) show 
100k events generated~\cite{vheep} using the {\sc Ariadne} Monte Carlo programme with the CTEQ2L proton PDFs for $Q^2>1$\,GeV$^2$, 
$x>10^{-7}$ and 
$W^2> 5$\,GeV$^2$. (b,d) show 500k events generated using either {\sc Rapgap 3.3} (orange line) or {\sc Herwig 7.0} (blue line) Monte Carlo 
programmes with the NNPDF 3.0 LO proton PDFs for $Q^2>1$\,GeV$^2$.}
\label{fig:q2-y}
\end{figure}

Within the Rivet framework, simulations of high energy $ep$ collisions were performed with the {\sc Rapgap}~\cite{rapgap} and 
{\sc Herwig}~\cite{herwig} Monte Carlo programmes.  These were run with more modern proton PDFs than used for {\sc Ariadne}, 
namely CT10~\cite{ct10}, MRST2004 FF4 LO~\cite{mrst2004} and NNPDF 3.0 LO~\cite{nnpdf}.  The programmes could also 
simulate events at very low Bjorken $x$, whereas the results from previous simulations with {\sc Ariadne} were restricted to a 
minimum Bjorken $x$ of $10^{-7}$.  The results of the simulations and 
comparison to previous results are shown in Fig.~\ref{fig:q2-y}.  The variables photon virtuality, $Q^2$, and inelasticity, $y$, are 
shown for {\sc Ariadne} (Fig.~\ref{fig:q2-y}(a,c)) and {\sc Rapgap 3.3} and {\sc Herwig 7.0} (Fig.~\ref{fig:q2-y}(b,d)).  The shapes of the 
three simulations are similar versus $Q^2$, most clearly shown for {\sc Rapgap 3.3} and {\sc Herwig 7.0}.  The shapes are also 
similar versus $y$, although a difference between {\sc Rapgap 3.3} and {\sc Herwig 7.0} can be seen.  

\begin{figure}
\begin{center}
\includegraphics[width=0.7\textwidth]{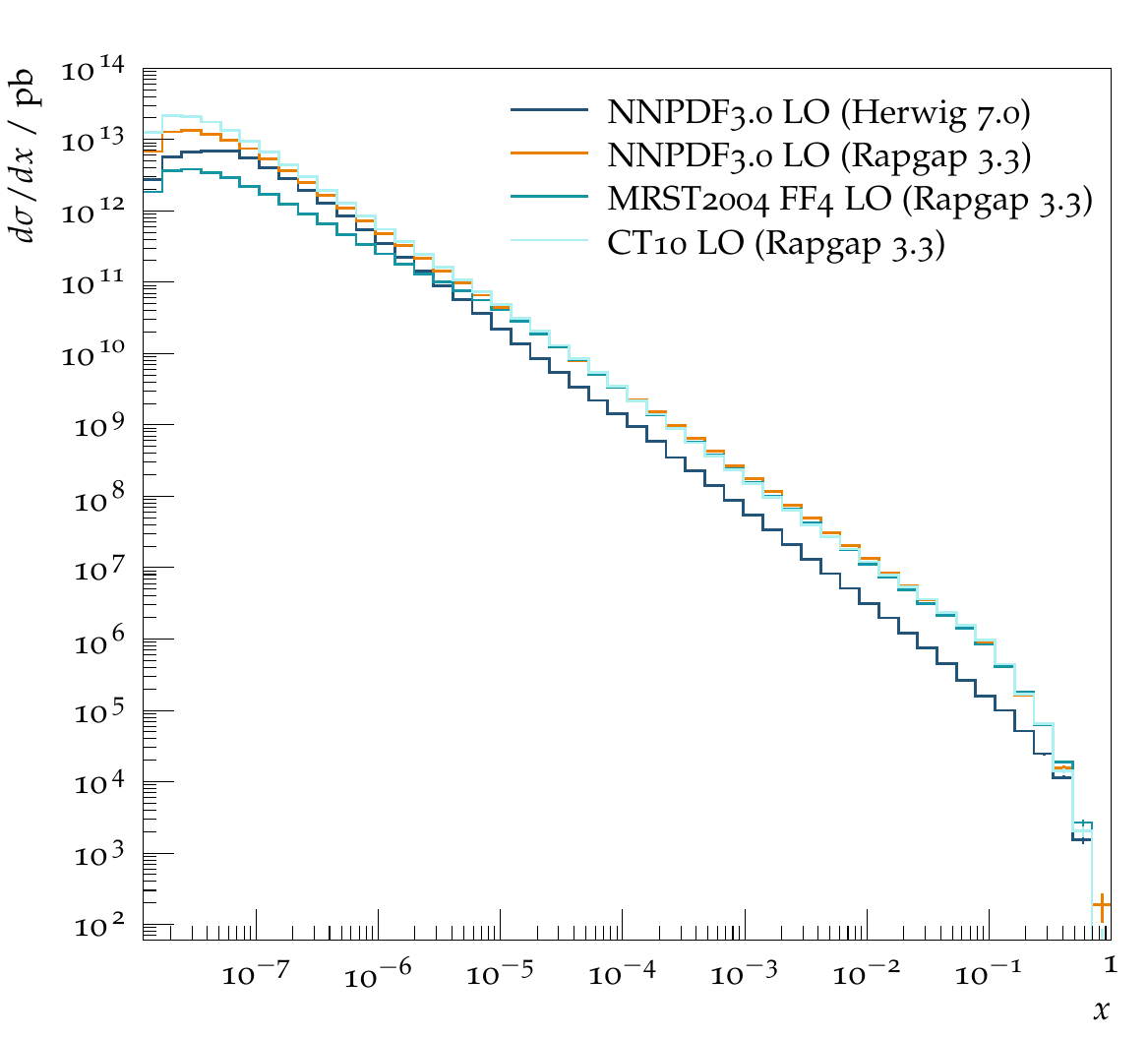}
\end{center}
\caption{Comparison of the Bjorken $x$ distribution when using either {\sc Rapgap 3.3} (orange line) or {\sc Herwig 7.0} (dark blue line) Monte Carlo 
programmes with the NNPDF 3.0 LO proton PDFs and also the {\sc Rapgap 3.3} Monte Carlo programme with either MRST2004 FF4 LO 
or CT10 LO proton PDFs.}
\label{fig:x}
\end{figure}

A further comparison 
between {\sc Rapgap 3.3} and {\sc Herwig 7.0} and between the central values of different proton PDFs, CT10 LO, MRST2004 FF4 LO and NNPDF3.0 LO, 
using {\sc Rapgap 3.3} is shown for the cross section versus Bjorken $x$ in Fig.~\ref{fig:x}.  The predictions from {\sc Rapgap 3.3} 
with different proton PDFs are very similar for Bjorken $x > 10^{-5}$, where they are well constrained by current data.  The PDFs 
deviate from each other to lower Bjorken $x$, where they are extrapolated under different assumptions.  Given the lack of data at 
very low Bjorken $x$, the true uncertainty is larger than the spread in these predictions, however, these predictions can be used as 
representative event simulations.  Differences between {\sc Herwig 7.0} and {\sc Rapgap 3.3}, using the same proton PDF, are observed 
over the full Bjorken $x$ range, even in the region well constrained by data.  This indicates differences in the simulations which should 
be investigated further.  It should be noted that the Monte Carlo simulations can be validated with HERA data within {\sc Rivet}, although 
the number of analyses from H1 and ZEUS number just a few at the time of writing.  It is hoped this will improve in the future and such 
differences between simulations can be understood.

\section{VHEeP workshop}
\label{sec:workshop}

A workshop~\cite{vheep-workshop}, "Prospects for a very high energy $ep$ and $eA$ collider", was held soon after the DIS 2017 workshop.  
The workshop covered two full days, held in the Max Planck Institute for Physics, Munich, with experts in the field invited to further explore 
the particle physics case for VHEeP.  The VHEeP baseline parameters with ranges which physics studies should consider are the following:

\begin{itemize}

\item collisions are nominally electron--proton, but electron--ion (e.g.\ electron--lead) collisions are foreseen;

\item there should be the possibility of accelerating positrons and having positron--proton collisions;

\item polarisation of the lepton beam should be possible and polarised protons should be considered;

\item the nominal energies are $E_e = 3$\,TeV and $E_p = 7$\,TeV, giving a centre-of-mass energy of $\sqrt{s} = 9.2$\,TeV;

\item the electron beam energy can be varied in the range 0.1 to 5\,TeV, giving $\sqrt{s} = 1.7 - 11.8$\,TeV;

\item the integrated luminosity is in the range 10 -- 1000\,pb$^{-1}$.

\end{itemize}

If different parameters are required, this will need to be considered with the constraints imposed by the acceleration scheme.

A brief summary of the workshop is given here and the interested reader is referred to the slides on the workshop 
web-page~\cite{vheep-workshop}.  The workshop started with an introduction to plasma wakefield acceleration and its application 
to higher energy physics (Caldwell), a report on the status of AWAKE (Muggli) and an introduction to VHEeP (Wing).  Of particular 
note were the presentations on QCD and hadronic cross sections (Bartels, Mueller, Schildknecht and Stodolsky) in which theoretical 
expectations show that saturation will be observed at VHEeP and will also be at a scale where QCD calculations are perturbative.  
Low-$x$ physics was related to other areas, including cosmic rays (Stasto) and black holes and gravity (Erdmenger), as well as being used 
as a testing-bed for new physics descriptions (Dvali and Kowalski).  The needs of polarisation and electron--ion physics were discussed 
(Aschenauer and M\"{a}ntysaari) as well as what has been learnt from the HERA data at low $x$ (Myronenko).  Finally, the status of 
Monte Carlo simulations for $ep$ and $eA$ physics was presented (Pl\"{a}tzer) as well as the general kinematics and challenges 
faced by the detector as well as the use of simulations (Keeble).


\section{Summary}
\label{sec:summary}

These proceedings have highlighted recent progress on studies for a very high energy electron--proton (VHEeP) collider since 
the proposal was published.  Simulations have been extended and can be used to investigate the physics potential and aid 
design of the detector and accelerator.  A workshop to further investigate the physics potential of electron--proton and electron--ion 
collisions at very high energies was held in June and was here briefly summarised.  It should also be investigated how VHEeP 
could fit into the future world particle physics programme and complement other similar projects investigating deep inelastic 
scattering.


\end{document}